\documentclass[aps,prl,twocolumn]{revtex4}

\usepackage{graphicx,latexsym,color}

\def\be{\begin{equation}}
\def\ee{\end{equation}}
\def\bea{\begin{eqnarray}}
\def\eea{\end{eqnarray}}
\def\bi{\begin{itemize}}
\def\ei{\end{itemize}}
\begin{document}
\title{Dynamics of Bloch Oscillations in Disordered Lattice Potentials}
\author{T. Schulte$^{1,2}$, S. Drenkelforth$^1$, G. Kleine B\"uning$^1$,
  W. Ertmer$^1$, J. Arlt$^{1,*}$, M. Lewenstein$^{2,3}$, and L. Santos$^4$}
\affiliation{$^1$ Institut f\"ur Quantenoptik, Leibniz Universit\"at Hannover,
  Welfengarten 1, D-30167 Hannover, Germany}
\affiliation{$^2$ ICFO - Institut de Ci\`encies Fot\`oniques,
E-08860 Castelldefels (Barcelona), Spain}
\affiliation{$^3$ ICREA - Instituci\`o Catalana  de Recerca i Estudis Avan\c cats, E-08010 Barcelona, Spain}
\affiliation{$^4$ Institut f\"ur Theoretische Physik, Leibniz Universit\"at Hannover, D-30167 Hannover, Germany}
\begin{abstract}
We present a detailed analysis of the
dynamics of Bloch oscillations of Bose-Einstein condensates
in disordered lattice potentials. Due to the disorder and the interparticle
interactions
these oscillations undergo a dephasing, reflected in a damping of the center
of mass oscillations, which should be observable under realistic
experimental conditions. The interplay between interactions and disorder
is far from trivial, ranging from an interaction-enhanced damping
due to modulational instability for strong interactions, to an
interaction-reduced damping due to a
\textit{dynamical screening} of the disorder potential.
\end{abstract}
\maketitle

Bloch oscillations (BOs) constitute one of the most fundamental quantum
phenomena for particles in periodic potentials.
Under the influence of a constant force, particles in such potentials undergo an oscillatory
motion instead of being linearly accelerated \cite{FKBuch,BlochZener}.
Although BOs are strongly linked to the dynamics of
electrons in solids, they have
not been observed in bulk crystalline materials so far, since
lattice imperfections such as defects and phonons
damp the coherent electronic motion before a single BO cycle is completed. The
first observation of BOs was achieved in so-called semiconductor super
lattices \cite{SCSLBO}, which exhibit much larger periodicities.
However, disorder also leads to a fast decay of the BOs in these systems.

On the contrary, ultra-cold gases in optical lattices provide
perfect periodic potentials, with neither defects nor phonons. As a
consequence, these systems open unprecented possibilities for the
detailed analysis of quantum transport in lattices, and in
particular for the observation of BOs with very long lifetimes
\cite{BOCG}. However, in spite of their perfect periodicity, optical
lattices allow for the controlled introduction of different types of
disorder \cite{Speckle, Bogdan, superlattice, impurity}. Recent
experiments have analyzed the effects of controlled disorder on the
properties of ultra-cold gases \cite{CollExc, ExpDyn, Wir,
Hamburg,BoseGlass,Modug}. These experiments have clearly shown that
ultra-cold gases are indeed very promising systems for the analysis
of the intriguing interplay between disorder and interparticle
interactions.

This paper is devoted to the analysis of this interplay in the BO
dynamics of Bose-Einstein condensates (BECs) in tilted optical
lattices. The dynamics of BOs is usually analyzed in terms of the
Wannier-Stark (WS) energy ladder in such systems. Any disorder
introduces an unequal spacing in this ladder and thus leads to a
damping of the BOs. In addition, the BOs are non-trivially modified
by the interactions in the system. Strong interactions enforce
damping, due to dynamical instability. However, weak interactions
can cause a \textit{dynamical screening} of the disorder potential,
prolonging the lifetime of the BOs \cite{Modugnopaper}. Although
quasi one-dimensional systems allow for a qualitative understanding
of the physics involved, in typical experimental conditions radial
excitations play a non-negligible role. In this paper we therefore
first develop a qualitative understanding of the dynamics of BOs in
the 1D case and then extend our analysis to the experimentally
relevant 3D case. Finally, we discuss a method for the observation
of damped BOs by analyzing the momentum distribution in
time-of-flight (TOF) measurements.

Consider the quasi-1D case, in which a BEC at low temperature
is so strongly confined by an harmonic trap of frequency $\omega_\perp$
in the $xy$ plane, that the chemical potential $\mu$ is much smaller then the transverse level spacing $\mu<<\hbar\omega_\perp$. Under these conditions, the $1D$ dynamics of the condensate wave
function $\Phi$ along the $z$-axis, is given by the Gross-Pitaevskii-equation
(GPE) \cite{StringariBuch}
\begin{equation}\nonumber
i \hbar \partial_{t} \Phi= \left[
  \frac{-\hbar^{2}}{2m}\frac{\partial^2}{\partial z^2}+V(z) + g
|\Phi|^{2}\right]\Phi,
\label{GPE}
\end{equation}
where $m$ denotes the atomic mass, and $g=4 \pi a_s \hbar \omega_{\perp}$
is the 1D coupling constant, with $a_s$ the $s$-wave scattering length.
$V(z)$ denotes an external potential of the form
\begin{equation}
V(z)=m\omega^{2}z^{2}/2+s\,E_{r}\sin^2(\bar kz)-Fz+V_{dis}(z).
\end{equation}
i.e. a superposition of an axial harmonic trap with frequency $\omega$,
a tilted potential with slope $F$, a disorder potential $V_{dis}(z)$, and
a lattice potential of periodicity $d=\pi/\bar k$,
and depth $s$ in units of the recoil energy $E_r=\hbar^2 \bar k^2/2m$.
The condensate ground state in the superimposed harmonic and
lattice potential serves as the initial state for our simulations of
the dynamics.

\begin{figure}[h]
\centering
\includegraphics*[width=8.5cm]{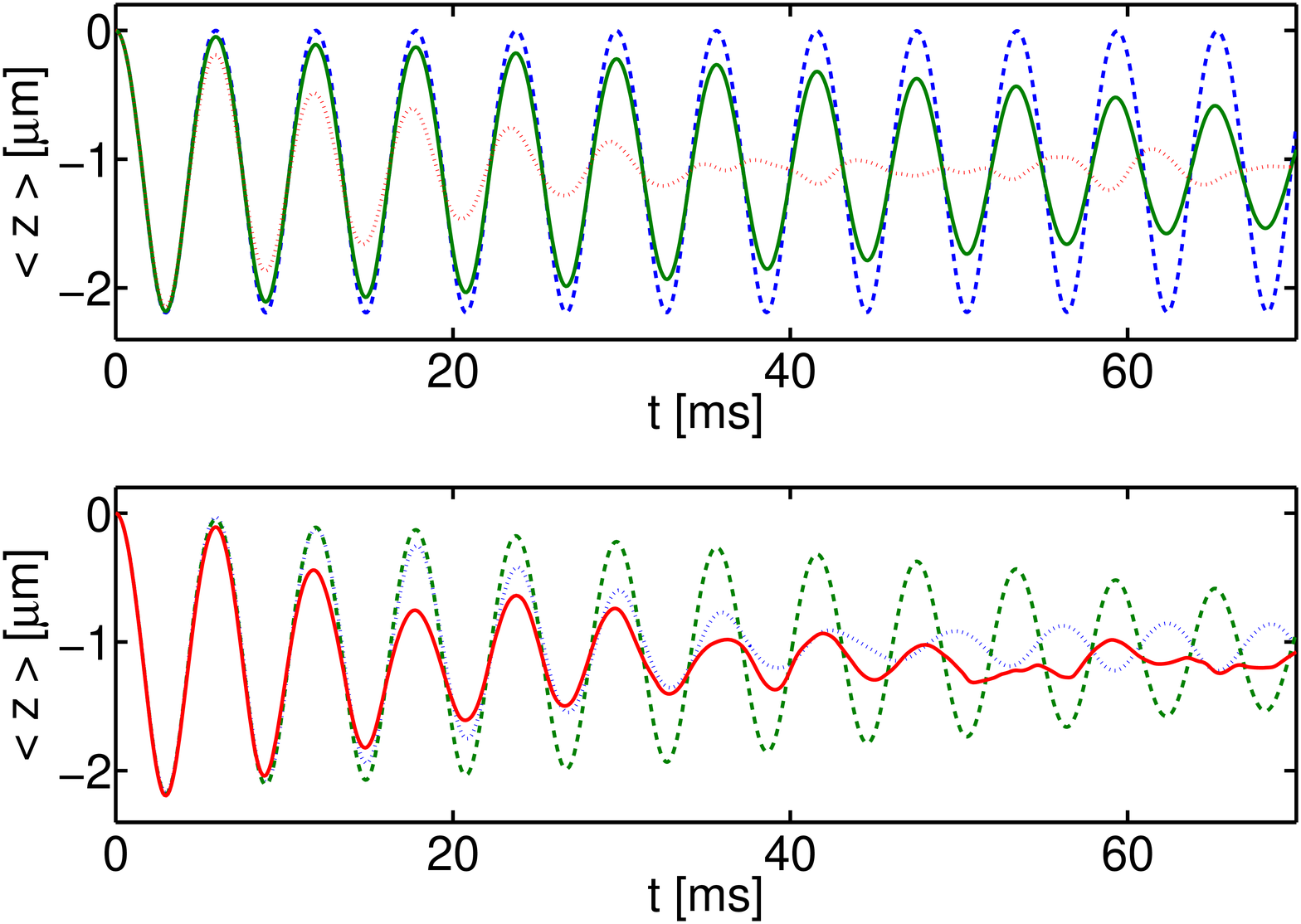}
\caption{Averaged position of a BEC undergoing BOs. Top frame: $N=350$ particles, $\omega=7\,Hz$, and disorder depths
$V_{\Delta}/E_r=0$ (dashed), $0.02$ (solid), and
$0.04$ (dotted). Bottom frame: $V_{\Delta}/E_r=0.02$ for
$N=1$, $\omega=3.5\,Hz$ (dotted), $N=350$, $\omega=7\,Hz$
(dashed) and $N=700$, $\omega=10\,Hz$ (solid).
The trap frequencies were adjusted to match the initial wavepacket
widths.}
\label{fig:1}
\end{figure}

Figure~\ref{fig:1} shows the averaged position
$\langle z(t)\rangle =\int dz\,\Phi^* z \Phi$ for a Rubidium condensate,
with $\omega_{\perp}=2\pi\times 200$ Hz, $d=412.5$ nm, $s=5$
and $Fd/E_r=0.05$, for different particle numbers $N$,
axial trap frequencies $\omega$, and disorder depths.
As disorder potential,
we consider Gaussian noise with correlation length $L\simeq 3.3\,\mu$m.
We define the disorder depth, $V_\Delta$
as twice the standard deviation
from its mean value, according to \cite{CollExc, Wir}.
Figure~\ref{fig:1} demonstrates that even for situations
where interactions are negligible ($N=1$)
disorder induces strong BO damping. The role of the interactions is
far from trivial, as shown in Fig.~\ref{fig:1}, since they may
enhance or significantly reduce the BO damping. We analyze this intriguing
physics in detail below.

\begin{figure}[h]
\centering
\includegraphics*[width=7cm]{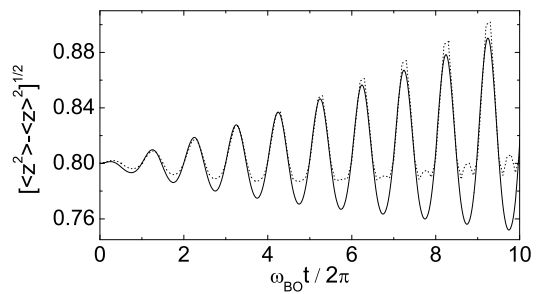}
\caption{
Condensate width in units of $z_{BO}$ obtained from direct simulation of
(\ref{GPE}) (dashed) and from the WS analysis discussed in the text (solid),
for $\hbar\omega_{BO}=0.08J$, $U/\hbar\omega_{BO}=3.7$, and
$\epsilon_n/\hbar\omega_{BO}=0.1\cos 0.16\pi n$.}
\label{fig:2}
\end{figure}

The BO damping can be understood from a simplified
analysis in the tight-binding regime. In this regime, we may project the
condensate wavefunction into the Wannier basis
of localized states $|n\rangle$, where $n$ labels the lattice sites.
$\Phi_n\equiv \langle n | \Phi\rangle$ is provided
by the discrete non-linear Schr\"odinger equation (DNLSE)\cite{DNLSE}
\begin{equation}
i\hbar\dot \Phi_n=\hat H_0\Phi_n+\epsilon_n\Phi_n+U|\Phi_n|^2\Phi_n,
\label{eq:DNLSE}
\end{equation}
where $\epsilon_n$ denotes the external (possibly disordered) potential
and $U$ the on-site interaction energy. $\hat H_0$
is the Hamiltonian for the tilted lattice in the absence of both
disorder and interactions,
$
\hat H_0 \Phi_n=-J\sum_n (\Phi_{n+1}+\Phi_{n-1})+F d n \Phi_{n}
$,
with hopping constant  $J$, and tilting potential $F d n$.
In absence of disorder and interactions, the particles perform BOs
with frequency $\omega_{BO}=F d /\hbar$, and
amplitude $z_{BO}=2Jd/\hbar\omega_{BO}$. The eigenstates of
$\hat H_0$ are the WS states \cite{Hartmann2004},
$|W_m\rangle=\sum_n J_{n-m}|n\rangle$, where
$J_n\equiv J_n(z_{BO})$ is the
Bessel function of first kind. The eigenenergies form the
well-known WS-ladder $E_m^0=m\hbar\omega_{BO}$.
We project the DNLSE in the $WS$ basis
$\psi_m\equiv\langle W_m|\Phi\rangle$
\begin{equation}
i\hbar \dot \psi_m=E_m^0 \psi_m + \sum_{m'} B_{mm'}\psi_{m'}
+
\sum_{m',s,s'}A_{m,m'}^{s,s'}\psi_{m'}^*\psi_{s'}\psi_{s},
\label{WS-eq1}
\end{equation}
where $B_{mm'}\equiv \sum_n \epsilon_n J_{n-m}J_{n-m'}$, and
$A_{m,m'}^{m'',m'''}\equiv U\sum_n J_{n-m}J_{n-m'}J_{n-m''}J_{n-m'''}$.
If $\hbar\omega_{BO}$ is much larger than other energy scales we can
neglect terms in (\ref{WS-eq1}) that introduce energy jumps larger
than $\hbar\omega_{BO}$, i.e. we can
employ rotating-wave-approximation (RWA) arguments.
In the RWA the disorder preserves
$\rho_m\equiv |\psi_m|^2$, just providing a shift $E_m^0+B_{mm}$.
Note, that on a longer time scale of several BOs
the RWA may fail and the disorder eventually leads to a transfer
of population between WS states. The interactions can, even in the RWA, lead to a transfer
of population between WS states. However, since
$\dot\rho_m\sim\rho_m^2$ whereas the phase $\varphi_m$ of $\psi_m$
evolves as $\dot\varphi_m\sim\rho_m$, we may consider $\rho_m$ as being constant
(at least at short time scales of few BOs) if the atomic wavepacket
is sufficiently broad \cite{Witthaut2005}. In that case, the energy of
the WS states becomes
\begin{equation}
E_m\simeq m\hbar\omega_{BO}+B_{mm}+2\sum_n \left (
\Gamma_n^e\rho_{m-2n}+\Gamma_n^0\rho_{m-2n-1}
\right ),
\label{Em}
\end{equation}
with $\Gamma_n^e\equiv A_{n,n}^{n,n}$, and
$\Gamma_n^0\equiv A_{n+1,n}^{n+1,n}$. Hence, the energies of the WS states are
not equidistantly spaced, the wavepacket undergoes dephasing, and as a consequence the
BOs are damped. In addition, the BEC width experiences
a breathing dynamics \cite{Witthaut2005}, as shown in Fig.~\ref{fig:2}.
This figure shows (at least during the first BOs) a good agreement
between our results from a direct
simulation according to equation (\ref{eq:DNLSE}) and those obtained
using WS states $\psi_m(t)=\sqrt{\rho_m(0)}e^{-iE_mt/\hbar}$,
where $E_m$ is given by (\ref{Em}).

Let us now discuss the intriguing role of the interactions on the BOs in
more detail. In the lower panel of Fig.~\ref{fig:1}, a stronger damping then in the single particle case can be observed for large
nonlinearity (see the curve for $N=700$). This interaction-induced damping
is related to the so-called dynamical instability
\cite{ModugnoInst,ChiaraBloch}. This instability occurs when the
quasi-momentum reaches the outer parts of the Brillouin zone and
small perturbations of the condensate wave function grow
exponentially in time \cite{ModugnoInst}. This mechanism becomes
predominant with growing nonlinearity, strongly damping the
BOs.

On the contrary, the BO damping may be significantly reduced compared to the single particle case for weak nonlinearity (see the curve for $N=350$). This effect is caused by an interaction-induced
{\it dynamical screening of the disorder} \cite{statscreen} and can be qualitatively better understood with an alternative
semi-classical description of the BOs. In the regime of weak nonlinearity,
we can assume that the dynamics occurs within the lowest Bloch band, and
that the dynamical instability is irrelevant on the time-scales considered.
Let us denote the exact mean field potential
obtained by solving the GPE~(\ref{GPE})
by $V_{mf}(z,t)= g|\Phi(z,t)|^2$, and consider the effective single particle
problem, for a particle in the lowest Bloch band under the influence of
$V_{mf}(z,t)$, the tilting force $F$ and the disorder $V_{dis}(z)$. The single
particle Hamiltonian is given by $H_{eff}= -2J \cos(\hat k d)-F\hat z +V_{dis}(\hat z)+V_{mf}(\hat z,t).$
In the quasi-momentum picture, $\hat k\rightarrow k$, and
$\hat z=i\partial/\partial k$. We assume that the bare amplitude of the BOs,
$z_{BO}$ is much smaller then the spatial spread of the initial
wave function.  This
in turn implies a very narrow initial momentum distribution, centered
at $k_0$, so that, in the absence of disorder and nonlinearity (zero-order solution),
$\hat z(t)\simeq \hat z(0) -z_{BO}\cos [ (Ft/\hbar+k_0)d ]$. The full
Heisenberg equation for $\hat k$ reads
$d\hat k/dt= F-(\partial V_{dis}/\partial z)(\hat z)-(\partial V_{mf}/\partial
z)(\hat z,t)$.
We solve it pertubatively by inserting the zero order solution. Again,
assuming a sharp initial momentum distribution we obtain
$\hat z(t)\simeq \hat z(0)
-z_{BO}\cos \{ \frac{Ftd}{\hbar}+k_0d-\frac{d}{\hbar}\int_0^t\ dt'
[(\partial V_{dis}/\partial z)(\hat z)-(\partial V_{mf}/\partial z)(\hat
z,t')] \}$.
In order to calucalte the dephasing
rate the latter expression  has to be averaged over the initial spread of
$\hat z(0)$. A reasonable estimate of the rate is $\gamma^2\simeq
\frac{d^2}{\hbar^2 t^2} \langle
\{
\int_0^t\ dt' [(\partial V_{dis}/\partial z)(\hat z(0))+
(\partial V_{mf}/\partial z)(\hat z(0),t')]
\} ^2
\rangle $.
Note that, when acting alone, both disorder and nonlinearity lead to the
damping of the BOs. However, when acting together, they may
compensate each other if the product of the time
averaged forces due to disorder and nonlinearity averaged over
$\hat z(0)$ is negative, qualitatively explaining the dynamical screening of
the disorder observed in Fig.~\ref{fig:1}.

\begin{figure}[h]
\centering
\includegraphics*[width=7cm]{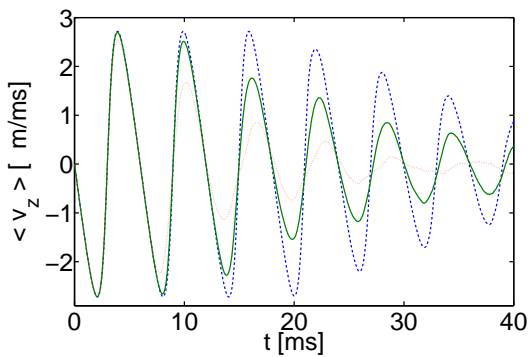}
\caption{Expectation value of the axial velocity for disorder
depths $0$ (dashed), $0.02\,E_r$ (solid) and $0.06\,E_r$ (dotted), for the 3D example discussed in the text.}
\label{fig:4}
\end{figure}

\begin{figure}[h]
\centering
\includegraphics*[width=6.5cm]{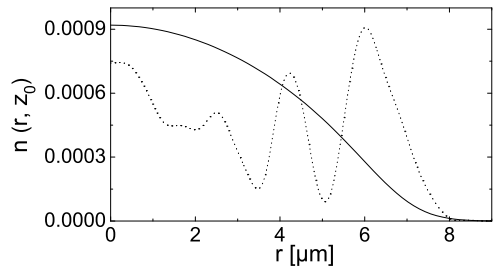}
\caption{Radial density profiles at the axial position $z_0$
of maximal density, at $t=0$ (solid), $16$ms (dotted)
for the case discussed in the text.}
\label{fig:5}
\end{figure}

In the previous discussion we have constrained our analysis to the
somewhat simplified quasi-1D regime, where the $xy$ dynamics
is absent. However, typical experiments are not performed in this regime,
and transversal excitations of the condensates may signficantly
alter the BO dynamics. Hence a quantitative description
of the BOs demands a three-dimensional GPE simulation.
We maintain a cylindrical trap, and hence we simplify
our 3D calculations by assuming cylindrical symmetry of the
wavefunction around the $z$-axis. Since experiments typically
detect BOs by observing the velocity distribution, we analyze
the expectation value of the axial velocity
$<v_z>=\int d^3r \,\Phi^* \frac{\hbar}{i m}
\partial_{z} \,\Phi$.
Figure~\ref{fig:4} shows the case
of $N=5\times 10^4$ particles, for the same lattice tilting
considered before, but for trap frequencies
$\omega_{z}=2\pi\times 14$Hz, $\omega_{\perp}=2\pi\times 35$Hz, lattice depth $s=2$,
and a disorder potential of depth $V_\Delta=0.02\,E_r$ and
correlation length $L\simeq10\mu$m.
These parameters are well attainable experimentally \cite{Wir}, and hence our simulations show that the damping of BOs can be studied under realistic conditions. For the large number of atoms considered here, damped BOs are
observed even in the absence of disorder due to a fast damping by the dynamical
instability. Under these conditions our 3D results
clearly deviate from the expected quasi-1D results, due to
the development of a complex radial dynamics shown in Fig.~\ref{fig:5}, due to instability of the
radial excitations.  Hence, although the dynamical
instability is also encountered in 1D, the full computation of the
dynamics does require the careful
consideration of the radial degree of freedom.

\begin{figure}[h]
\centering
\includegraphics*[width=7cm]{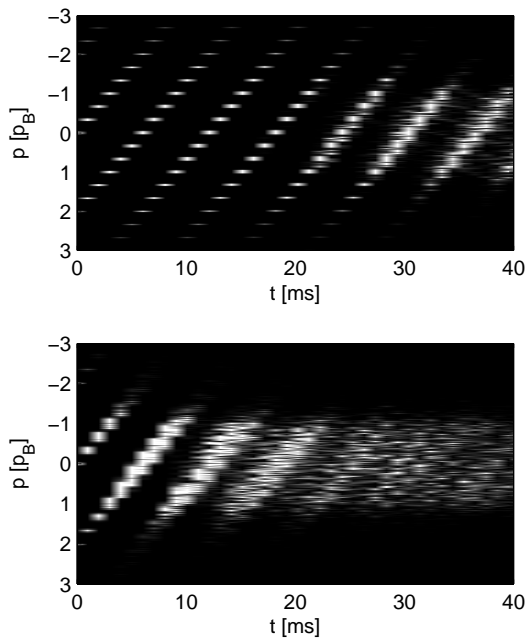}
\caption{Gray-scale plot of the momentum distribution $n(p_{\perp}=0,p_z)$,
  calculated every $1\,ms$ during the BOs. The upper panel shows
the non-disordered case, the lower panel shows the result for a
disorder depth of $0.06\,E_r$. The pictures correspond to the 3D
configuration of Figs.~\ref{fig:4} and \ref{fig:5}.}
\label{fig:6}
\end{figure}

Finally, we discuss a method for the experimental observation of damped BOs. A direct measurement
of the damping in position space constitutes a very difficult task,
since the oscillation amplitude is typically
too small for an in-situ measurement. TOF measurements provide a much better
method. Due to the
very fast decrease of the mean-field energy during the TOF, the expanded
density distribution provides an approximative image of the momentum distribution $n(\vec p)$ of the sample at the moment of release. Figure~\ref{fig:6} shows the momentum distribution at various stages during the BOs. In the absence of
disorder, the quasi-momentum $q$ scans the Brillouin-zone
due to the acceleration introduced by the tilting force
($\dot q=F/\hbar$) \cite{Hartmann2004}. Consequently, the
population of the different momentum components changes, resulting
in a coherent oscillation of the mean momentum. The spectrum
consists of several sharp peaks, which are separated from each
other by the lattice momentum $p_B=2\hbar \bar k$. Hence, the density distribution after an expansion time $\tau$ consists of
several small clouds, well separated from each other by
$\Delta z=2 p_B \tau/m$. This picture changes
when disorder is introduced to the system. The initial sharp
momentum components are progressively broadened, eventually
reaching an irregular occupation of momenta. This broadening
originates from the irregular energy
spacing of the WS states and leads to
a broadened density distribution after TOF. Note, that on long time scales, a significant broadening of the
momentum components is also introduced by the interactions in the absence of
disorder, as shown in Fig.~\ref{fig:6}. However, a very clear distinction between the ordered and disorderd case is possible on time scales of up to a few BO periods. Moreover, our 3D simulations
of the expansion dynamics show that it is possible to observe a clear
separation between the center of mass position for the ordered
and disordered lattices under realistic conditions, due to the differences in the expected value
of the momentum in both cases (Fig.~\ref{fig:4}). Also
note that Fig.~\ref{fig:6} shows the axial momentum distribution
for zero transversal momentum, which is significantly reduced
during the time evolution. This reflects the previously mentioned
excitation of transverse modes.

In summary, both disorder and interactions separately lead
to BO damping. However, when acting simultaneously the
interactions may partially screen the disorder, leading to a reduction
of the BO damping. We have shown that the interplay of disorder and interactions may be observed
under realistic experimental conditions by monitoring
the evolution of the momentum distribution of the system.
Although we were mainly interested in disordered lattices,
similar results may be obtained for lattices in the presence
of inhomogeneous forces, as e.g. spatially inhomogeneous Casimir-Polder
forces close to surfaces. It has recently been proposed that
the frequency shift of BOs of lattice fermions can be an
excellent way of measuring such tiny forces \cite{Carusotto}.
The monitoring of BO damping and the sample width may provide
an excellent way of proving the inhomogeneity of these forces.

We thank L. Sanchez-Palencia for fruitful discussions. We
acknowledge support from the programs QUDEDIS (grants 1365,1551) and Euroquam FERMIX of the European Science Foundation (ESF). This work was supported by the Deutsche
Forschungsgemeinschaft (SFB 407, SFB-TR21, SPP1116), Spanish MEC (FIS 2005-04627, Consolider Ingenio 2010 QOIT).


\begin{thebibliography}{99}
\bibitem{Modugnopaper} A similar dynamical screening may explain some features observed
 in dipole oscillations in Ref.\cite{Modug}.
\end{thebibliography}
\end{document}